\documentclass{article}
\usepackage{amsmath}
\usepackage{graphicx}
\usepackage{tabls}
\usepackage{afterpage}
\usepackage{epsf}

\usepackage{amsfonts}

\newcommand{\BE}{\begin{equation}}
\newcommand{\EE}{\end{equation}}

\newcommand{\cI}{\mathcal{I}}
\newcommand{\cJ}{\mathcal{J}}

\newcommand{\cW}{\mathcal{W}}

\newcommand{\mR}{\mathbb{R}}
\newcommand{\e}{\mathrm{e}}

\newcommand{\bk}[1]{{\langle #1 \rangle}}
\newcommand{\abs}[1]{{\vert {#1} \vert}}

\newcommand{\pt}{\partial}

\title{Adding decoherence to the Wigner equation}

\author{L.~Barletti, G.~Frosali, G.~Frosali} 
\date{\normalsize Dipartimento di Matematica e Informatica ``U. Dini'' \\
Universit\`a degli Studi di Firenze \\
Viale Morgagni 67/A, I-50134 Firenze, Italia}

\begin{document}

\normalsize

%      Headers and Footers

%\afterpage{%
%\fancyhf{}%
%\fancyhead[CE]{              
%{\scriptsize \authorHead}}                                                
%\fancyhead[CO]{               
%{\scriptsize \shortTitle}}                  
%\rfoot{\thepage/\totalpages{}}%
%
%%     Change this if your installation causes the document to be shifted downward
%%\setlength{\topmargin}{-20pt}
%}
%
%\pagestyle{fancy}
 
\maketitle

%
% SET RAGGED RIGHT MARGIN
%
%\raggedright

%\setlength{\baselineskip}{14pt}
\normalsize

\begin{abstract}
Starting from the detailed description of the single-collision decoherence mechanism proposed by Adami, Hauray and Negulescu in Ref.\ \cite{AHN}, we derive
a Wigner equation endowed with a decoherence term of a fairly general form.
This equation is shown to contain well known decoherence models, such as the Wigner-Fokker-Planck equation, as particular cases. 
The effect of the decoherence mechanism on the dynamics of the macroscopic moments (density, current, energy) is illustrated by deriving the corresponding set of
balance laws.
The issue of large-time asymptotics of our model is addressed in the particular, although physically relevant, case of gaussian solutions. 
It is shown that the addition of a Caldeira-Legget friction term provides the asymptotic behaviour that one expects on the basis of physical considerations.
\end{abstract}

\section{Introduction}

Decoherence is the process of loss of quantum coherence \cite{JZ,Zurek}.
As such, it governs the transition from quantum to classical behaviour and it shapes our actual perception of the world.
The theoretical and experimental study of decoherence processes is not only important for our understanding of fundamental physics, 
but it is also crucial for those technological applications, such as quantum computers and spintronics \cite{Zutic02}, where quantum coherence 
must be preserved as long as possible.
In this contribution we intend to present a model  of dynamical quantum decoherence within the Wigner (phase-space) formulation of quantum mechanics
\cite{BUMI03,Jungel,Wigner32,Tatarskii83,ZachosEtAl05}.
In fact, due to its striking analogies with  classical mechanics, the formulation of quantum mechanics in terms of Wigner functions is particularly 
suited to illustrate the quantum-to-classical regime transition. 
Of course, this approach is not new and several important papers on this subject resort to (or, at least, mention) the Wigner formalism 
(see e.g.\ Refs.\ \cite{ALMS,CL,JZ,Schwaha13,Wi})
The novelty of the present paper is that we start from a decoherence model which is fairly general whenever the environment  is viewed as a 
``gas'' of particles  of asymptotically small mass, with respect to the ``heavy'' particle undergoing decoherence.
This model has been rigorously derived from the laws of quantum mechanics  in Refs.\ \cite{AdamiEtAl04,AHN},
and, to this extent, our description can be considered as arising ``from first  principles''.
Indeed, as we shall see, other models (above all the Wigner-Fokker-Planck equation) can be recovered as particular cases of the general mechanism introduced here.
\par
The content of the present paper is the following.
Here below we briefly present the single-collision decoherence model analyzed in  Refs.\ \cite{AdamiEtAl04,AHN}.
Then, in next section, we consider the case of many collisions, randomly distributed in time, and obtain the corresponding ``mean field'' limit model,
which is then translated into the Wigner framework.
In Section \ref{S3}, the Wigner equation with decoherence obtained in this way is shown to be strictly related to other models of decoherence, 
such as the Wigner-Fokker-Planck equation  \cite{ALMS,Arnold,DH,JZ} and the Jacoboni-Bordone Wigner function with finite coherence length \cite{JB}.
In Section \ref{S4} we study the influence of the decoherence mechanism on the dynamics of macroscopic quantities, namely density, current and energy.
Section \ref{S5} is devoted to the issue of long-time asymptotics: the numerical investigation of a simple situation (i.e.\ the case of gaussian solutions) suggests 
that the correct long-time behaviour requires the addition of a Caldeira-Legget ``quantum friction term'' \cite{CL}.
Finally, in Section \ref{S6} we draw some conclusions and discuss future perspectives.
\par
\medskip
The quantum dynamical decoherence of a heavy particle interacting with a single light particle is analysed in Refs.~\cite{AdamiEtAl04,AHN}. 
The main result of such analysis is as follows.
Let $\rho(X,Y,t)$ be the reduced density matrix of the heavy particle (the degrees of freedom of the light particle are traced out).
Then, in the limit of large heavy-to-light mass ratio, the interaction is concentrated in a 
single instant of time (say, $t = 0$) and  has the form of the ``instantaneous'' transformation
$$
\rho(X,Y,0) \longmapsto \cI(X,Y)\rho(X,Y,0),
$$
where $\cI(X,Y)$ is a ``collision factor'', depending on the details of the interaction.
Elsewhere, $\rho(X,Y,t)$ evolves freely (up to possible external potentials $V$).
In the one-dimensional case, this single-interaction decoherence mechanism model is therefore given by the von Neumann equation
with a modified initial datum:
\begin{equation}
\label{AHN}
\left\{
\begin{aligned}
 & i\hbar\frac{\pt \rho}{\pt t} = - \frac{\hbar^2}{2m}\left(\frac{\pt^2  \rho}{\pt X}  - \frac{\pt^2  \rho}{\pt Y}\right) + \left[V(X) - V(Y) \right]\rho,
 \\[12pt]
 & \rho(X,Y,0) = \cI(X,Y)\rho_0(X,Y),
\end{aligned}
\right.
\end{equation}
where $\rho_0(X,Y)$ is the pre-interaction density matrix.
The form of the collision factor $\cI$ is completely characterised in the one-dimensional case \cite{AHN} and is given by 
\begin{equation}
\label{Idef}
\cI(X,Y) = 1-\Lambda(X-Y) + i\Gamma(X) - i\Gamma(Y),
\end{equation}
with
\begin{align}
\label{LambdaDef}
&\Lambda(X) = \int_\mR (1- \e^{2ikX}) \abs{r(k)}^2  \abs{\hat\chi(k)}^2 dk,
\\[6pt]
\label{GammaDef}
&\Gamma(X) =  \int_\mR \e^{2ikX} \overline{r(-k)}\, t(k)\, \overline{\hat\chi(-k)}\, \hat\chi(k)\,dk
\end{align}
where $r$ and  $t$ are the scattering coefficients of the interaction, and $\hat\chi$ is the Fourier transform of the 
light-particle wave function $\chi$.
A particularly simple form can be obtained by assuming that
\begin{itemize}
\item[1)]
 $\chi$ is a gaussian wave-packet with average momentum $p_0 = \hbar k_0$ and position variance $\sigma^2$;
\item[2)]
  $p_0$ is large with respect to the momentum spread $\hbar\sigma^{-1}$;
\item[2)]
$\sigma^{-1}$ is small compared to the scale at which $\abs{r(k)}^2$ varies.
\end{itemize}
In this case, as shown in Ref.\ \cite{AHN}, one can make the following approximations:
\begin{equation}
\label{Approx}
\Lambda(X) \approx \abs{r(k_0)}^2 \left( 1- \e^{2ik_0X - \frac{X^2}{2\sigma^2}}\right),
\qquad
\Gamma(X) \approx 0.
\end{equation}
Such approximation, providing simple and explicit expressions,  will be helpful in the following.
\section{A Wigner equation with decoherence}
\label{S2}
We now consider a quantum particle undergoing random collisions with a gas of much lighter particles, each collision being 
described by the single-interaction model introduced above.
Let $\e^{tA}$ denote the unitary evolution group associated to the von Neumann equation 
$$
 i\hbar\frac{\pt \rho}{\pt t} = - \frac{\hbar^2}{2m}\left(\frac{\pt^2  \rho}{\pt X}  - \frac{\pt^2  \rho}{\pt Y}\right) + \left[V(X) - V(Y) \right]\rho,
$$
so that the solution to this equation with a generic initial datum $\rho_0$ is expressed (omitting the variables $X$ and $Y$) as 
$$
  \rho(t) = \e^{tA}\rho_0.
$$
Let $\nu$ be the collision probability per unit time, 
and let $\Delta t$ be a time-interval small enough to neglect the probability of having more than one collision inside it.
The random dynamics of the heavy particle  can be described by a density-matrix valued stochastic process $R(t)$ such that
$$
  R(t + \Delta t) = \left\{
  \begin{aligned}
   &\e^{\Delta t A} R(t), & &\text{with probability $1-\nu\Delta t$,}
   \\
   &\e^{(\Delta t - s)A} \cI \e^{sA} R(t), & &\text{with probability $\nu\Delta t$,}
  \end{aligned}
 \right.
$$
for some collisional time $s \in [0, \Delta t]$.
If now $\rho(t) = \bk{R(t)}$ is the expected value of $R(t)$, we clearly have 
$$
  \rho(t + \Delta t) = (1-\nu\Delta t)\e^{\Delta t A} \rho(t)  + \nu\Delta t \, \e^{(\Delta t - s)A} \cI \e^{sA} \rho(t)
$$
and then
$$
\frac{ \rho(t + \Delta t) - \e^{\Delta t A} \rho(t) }{\Delta t}  =  - \nu \e^{\Delta t A} \rho(t) + \nu \e^{(\Delta t - s)A} \cI \e^{sA}\rho(t).
$$
By using the fundamental property of the evolution group
$$
   \e^{\Delta t A}  = \int_0^{\Delta t} A \e^{\tau A} \, d\tau + I,
$$
we arrive at
$$
\frac{ \rho(t + \Delta t) - \rho(t) }{\Delta t} = \frac{1}{\Delta t} \int_0^{\Delta t} A \e^{\tau A} \rho(t) \,d\tau
- \nu \e^{\Delta t A} \rho(t) + \nu \e^{(\Delta t - s)A} \cI \e^{sA} \rho(t)
$$
and, taking the limit $\Delta t \to 0$ and recalling that $s \in [0, \Delta t]$, we obtain
$$
\frac{d\rho(t)}{dt} = A\rho(t) - \nu \rho(t) + \nu \cI \rho(t).
$$
By explicitly writing down this differential equation, and putting $\tau := \nu^{-1}$, we get
\begin{equation}
\label{VNEdeco}
 i\hbar\frac{\pt \rho}{\pt t} + \frac{\hbar^2}{2m}\left(\frac{\pt^2 \rho}{\pt X^2}  - \frac{\pt^2 \rho}{\pt Y^2}\right) + \left[V(X) - V(Y) \right]\rho
 =  \frac{i\hbar}{\tau}\left(\cI\rho- \rho\right),
\end{equation}
which is the von Neumann equation with a collisional term representing decoherence. 
The above formal derivation can be of course made rigorous by a suitable analysis. 
The rigorous derivation of Eq.\ \eqref{VNEdeco}, assuming the approximation \eqref{Approx}, is contained in Ref.\ \cite{HG}.
\par
Let us adopt a phase-space description in terms of the Wigner function \cite{BUMI03,Wigner32,ZachosEtAl05}, i.e.\ the Wigner transform $w = \cW\rho$ 
of the density matrix, where the Wigner transformation $\cW$ is defined as
\begin{equation}
\label{Wdef}
  \left(\cW\rho\right)(x,p,t) = \frac{1}{2\pi\hbar} \int_\mR \rho\left(x + \frac{\xi}{2},  x - \frac{\xi}{2},t \right) \e^{-i\xi p /  \hbar}\, d\xi.
\end{equation}
If we Wigner-transform the von Neumann equation \eqref{VNEdeco}, by using the property
\begin{equation}
\label{convprop}
  \cW(\rho_1\rho_2) = \cW(\rho_1)\ast\cW(\rho_2),
\end{equation}
we arrive at the following Wigner equation
\begin{equation}
\label{WE0}
   \frac{\pt w}{\pt t} + \frac{p}{m} \frac{\pt w}{\pt x} + \Theta[V] w = \frac{(\cW\cI)\ast w - w}{\tau},
\end{equation}
where $\cW\cI$ is the Wigner transform of the collision factor, $\ast$ denotes the  convolution with respect to the momentum variable $p$
and $\Theta[V]$ is the usual pseudo-differential operator
\begin{multline}
\label{Thetadef}
  \left(\Theta[V] w\right)(x,p,t) = \frac{i}{2\pi\hbar^2}
    \int_\mR \int_\mR \left[V\left(x+\frac{\xi}{2}\right) - V\left(x-\frac{\xi}{2}\right)\right] \e^{\frac{i}{\hbar}\xi(p'-p)} w(x,p',t)\,d\xi\,dp'
 \\[6pt]
    = - \sum_{k=0}^\infty (-1)^k \left(\frac{\hbar}{2}\right)^{2k} 
    \Big(\frac{d}{d x}\Big)^{2k+1}V(x) \,   \Big(\frac{\pt}{\pt p}\Big)^{2k+1}w(x,p,t).
\end{multline}
Note that if $V$ is a quadratic potential energy, or if one takes  the semiclassical limit $\hbar \to 0$, then $\Theta[V]$ reduces to the classical 
force term of the Liouville equation, namely
\begin{equation}
\label{Thetapprox}
  \Theta[V]w = - V'\, \frac{\pt w}{\pt p}\,,
\end{equation}
where, of course, $-V'$ is the force.
By using \eqref{Idef}, \eqref{Wdef} and \eqref{Thetadef}, we can write
\begin{equation}
\label{altform}
  (\cW\cI) \ast w - w
  = -\gamma \ast  w  + \Theta[\hbar\Gamma] w
\end{equation}
where
\begin{equation}
\label{gammadef}
     \gamma(p) = (\cW \Lambda)(p) =  \frac{1}{2\pi\hbar} \int_\mR \Lambda(\xi) \e^{-i\xi p /  \hbar}\, d\xi
\end{equation}
is a function of $p$ alone, because $\Lambda$ is a function of the correlation variable $\xi = X-Y$.
Hence, the Wigner equation \eqref{WE0} takes the final form
\begin{equation}
\label{WE}
  \frac{\pt w}{\pt t} + \frac{p}{m} \frac{\pt w}{\pt x} + \Theta\Big[V-\frac{\hbar}{\tau} \Gamma\Big]  w = - \frac{\gamma \ast w}{\tau}.
\end{equation}
Note that the term $-\frac{\hbar}{\tau} \Gamma$ is equivalent to a potential energy and, therefore, it contributes to the unitary evolution and not to the decoherence.
In the particular case of the peaked-gaussian approximation \eqref{Approx} we easily obtain 
\begin{equation}
\label{gaussform}
  - \frac{\gamma \ast w}{\tau}  \approx 
  \frac{\abs{r(k_0)}^2}{\tau} \left[\frac{\sigma}{\hbar\sqrt{2\pi}} \int_\mR
  \e^{-\frac{\sigma^2}{2\hbar^2}(p - p' - 2\hbar k_0)^2} w(x,p',t)\,dp' - w(x,p,t) \right].
\end{equation}
\par
With respect to the standard Wigner equation, Eq.\ \eqref{WE}  contains a decoherence mechanism which is represented by the 
right-hand side. 
Such equation is our basic model of dynamical quantum decoherence.
\par
The physical interpretation of Eq.\ \eqref{WE} is given as follows. 
The typical $\cI(X,Y)$ is a decaying function of the correlation distance $\abs{X-Y}$ (see Ref.\ \cite{AHN}). 
It means that  the decoherence process
$$
\rho(X,Y) \longmapsto \cI(X,Y)\rho(X,Y),
$$
results in a loss of spatial correlation.
Switching to the Wigner picture basically means performing a Fourier transform with respect to the correlation variable $\xi = X-Y$
and then the multiplication $ \cI(X,Y)\rho(X,Y)$ becomes a convolution with respect to the Fourier variable $p$.
Hence, the loss of spatial correlation corresponds to a smoothing out of $w(x,p,t)$ along the $p$ direction.
In particular, from Eqs.\ \eqref{Approx} and \eqref{gaussform}  we can see that, in the peaked-gaussian approximation, the position spread $\sigma$ of the light 
particle determines the reduction scale of the coherence length and, correspondingly, the momentum spread $\hbar\sigma^{-1}$ determines
the smoothing scale of the Wigner function. 
Roughly speaking, this mechanism attenuates the oscillations of the Wigner function (that are typically on a scale of order $\hbar$ in phase space \cite{Tatarskii83}), 
thus making the Wigner function progressively lose its quantum character and become a classical object.
\section{Relationships with other models} 
\label{S3}
By expanding $1- \e^{2ikX} = -2ikX + 2k^2X^2 + \cdots$, we obtain from \eqref{LambdaDef}
\begin{equation}
\label{LambdaExpansion}
   \Lambda(X) =  -i X \Lambda_1 + X^2 \Lambda_2 + \cdots,
\end{equation}
where, in the general case,
\begin{equation}
\label{Lambdaj}
  \Lambda_1 = 2 \int_\mR k\, \abs{r(k)}^2  \, \abs{\hat\chi(k)}^2 dk, 
  \qquad 
  \Lambda_2 = 2 \int_\mR k^2 \, \abs{r(k)}^2 \, \abs{\hat\chi(k)}^2 dk,
\end{equation}
and, in the approximation \eqref{Approx},
$$
  \Lambda_1 = 2k_0 \abs{r(k_0)}^2, \qquad  \Lambda_2 =  \frac{\abs{r(k_0)}^2}{2\sigma^2}.
$$
Then, we see from \eqref{gammadef} and \eqref{LambdaExpansion} that
\begin{equation}
 \gamma \ast w \approx -\hbar\Lambda_1\frac{\pt w}{\pt p}   - \hbar^2\Lambda_2 \frac{\pt^2 w}{\pt p^2} ,
\end{equation}
and, if one also assumes $\Gamma = 0$, the following model is obtained from \eqref{WE} and \eqref{altform}:
\begin{equation}
\label{WE2}
   \frac{\pt w}{\pt t} + \frac{p}{m} \frac{\pt w}{\pt x} + \Theta[V] w 
   =   \frac{\hbar^2\Lambda_2}{\tau}\frac{\pt^2 w}{\pt p^2} + \frac{\hbar\Lambda_1}{\tau}\frac{\pt w}{\pt p} .
\end{equation}
The term $\frac{\hbar \Lambda_1}{\tau}\frac{\pt w}{\pt p}$ is just a momentum drift due to our assumption that all environment particles are identical, having in particular the same momentum.
This rather unphysical assumption can of course be relaxed by assuming that the light particle is chosen at random from a given population. 
In this case, $\Lambda_1$ survives if the light-particle distribution is asymmetric with respect to the momentum.
Otherwise, if  $\abs{r(k)}^2$ and $\abs{\hat\chi(k)}^2$ are even functions of $k$ 
(or, simply, if $k_0 = 0$ in the approximation \eqref{Approx}), then $\Lambda_1 = 0$ and Eq.\ \eqref{WE2} 
reduces to the Wigner-Fokker-Planck equation 
\begin{equation}
\label{WFP}
   \frac{\pt w}{\pt t} + \frac{p}{m} \frac{\pt w}{\pt x} + \Theta[V] w =   \frac{\hbar^2\Lambda_2}{\tau} \frac{\pt^2 w}{\pt p^2},
\end{equation}
which is a largely used model of decoherence \cite{ALMS,Arnold,DH,JZ}.
\par
\medskip
By assuming $\Gamma = 0$ and 
$$
  1 - \Lambda(X) = \e^{-\abs{X}/\lambda},
$$
we obtain 
$$
   \cW(\cI\rho) =  \frac{1}{2\pi\hbar} \int_\mR \e^{-\abs{\xi}/\lambda}\,\rho\left(x + \frac{\xi}{2},  x - \frac{\xi}{2},t \right) \e^{-i\xi p /  \hbar}\, d\xi,
  =: w_\lambda(x,p,t).
$$
In this case, our model can be written
\begin{equation}
\label{WEJB}
   \frac{\pt w}{\pt t} + \frac{p}{m} \frac{\pt w}{\pt x} + \Theta[V] w = \frac{w_\lambda - w}{\tau},
\end{equation}
and can be interpreted as the dynamical analogous of the approach proposed by Jacoboni and Bordone in Ref.\ \cite{JB}, where 
a Wigner function with finite coherence length $\lambda$ is introduced,  which is exactly $w_\lambda$.
In fact, the decoherence mechanism contained in Eq.\ \eqref{WEJB} is clearly a relaxation of $w$ to $w_\lambda$ in a typical time $\tau$.
Recalling \eqref{convprop}, we can also write
$$
  w_\lambda(x,p,t) = (\cW\cI)\ast w = \frac{1}{\pi}\int_\mR\frac{\hbar/\lambda}{(\hbar/\lambda)^2 + (p-p')^2} \, w(x,p',t)\,dp',
$$
from which we see that the effect of the finite coherence length is a Lorentzian broadening of the Wigner function in momentum space,
as already remarked in Ref.\ \cite{JB}.
\par
Our approach allows a straightforward generalization of Eq.\ \eqref{WEJB}. 
In fact, it is enough to assume that the population of lighth particles has a non vanishing momentum $p_0$ to enrich Eq.\ \eqref{WEJB}
with the additional parameter $p_0$, namely 
$$
  w_{\lambda,p_0}(x,p,t)  = \frac{1}{\pi}\int_\mR\frac{\hbar/\lambda}{(\hbar/\lambda)^2 + (p-p_0-p')^2} \, w(x,p',t)\,dp',
$$
which embeds the momentum transfer from the environment to the particle undergoing decoherence.
\section{Balance laws}
\label{S4}
For the sake of conciseness, in what follows we assume $\Gamma = 0$. 
As we can see from Eq.\ \eqref{WE}, the general case with $\Gamma \not= 0$ is simply recovered by substituting 
$V$ with $V-\frac{\hbar}{\tau}\Gamma$.
\par
Balance laws can be deduced from the Wigner equation \eqref{WE} by taking suitable moments with respect to $p$.
In particular, we are interested in the following quantities:
\begin{equation}
\begin{aligned}
  &N(x,t) = \int_\mR w(x,p,t)\,dp,& &\text{(number density),} 
\\
  &J(x,t) = \frac{1}{m} \int_\mR p\,w(x,p,t)\,dp,& &\text{(current density),}
\\
 &E(x,t) = \frac{1}{2m} \int_\mR p^2 \,w(x,p,t)\,dp,&\quad & \text{(energy density).}
\end{aligned}
\end{equation}
In order to compute balance laws for $N$, $J$ and $E$, we need to take the corresponding moments of Eq.\ \eqref{WE} and,
in particular, we need the moments of $\Theta[V]w$ and $\gamma\ast w$.
By using the series expansion in Eq.\ \eqref{Thetadef}, it is readily seen that
\begin{equation}
\begin{aligned}
  &\int_\mR \left(\Theta[V] w\right)(x,p,t)\,dp = 0,
\\[6pt]
  &\frac{1}{m}\int_\mR p\left(\Theta[V] w \right)(x,p,t)\,dp = \frac{1}{m} V'(x) N(x,t),
\\[6pt]
 &\frac{1}{2m}\int_\mR p^2\left(\Theta[V] w \right)(x,p,t)\,dp = V'(x)\,J(x,t).
\end{aligned}
\end{equation}
Moreover, from \eqref{gammadef} and \eqref{LambdaDef} we obtain
\begin{equation}
  \int_\mR \gamma(p)\,dp = \Lambda(0) = 0,
\end{equation}
which means that the number of particles is conserved, and 
\begin{equation}
 \int_\mR p\,\gamma(p)\,dp = -i\hbar\Lambda'(0),
\qquad
 \int_\mR p^2\,\gamma(p)\,dp = -\hbar^2\Lambda''(0).
\end{equation}
With a little additional algebra we arrive at
\begin{equation}
\begin{aligned}
  &\frac{1}{m}\int_\mR p\left(\gamma\ast w\right) (x,p,t)\,dp = -\frac{i\hbar}{m} \Lambda'(0)\, N(x,t),
\\[6pt]
 &\frac{1}{2m}\int_\mR p^2\left(\gamma\ast w\right) (x,p,t)\,dp = -\frac{i\hbar}{m} \Lambda'(0)\, J(x,t) + \frac{\hbar^2}{2m} \Lambda''(0)\,N(x,t).
\end{aligned}
\end{equation}
From \eqref{LambdaExpansion} we see that $i\Lambda'(0) = \Lambda_1$ and $\Lambda''(0) = 2\Lambda_2$ (where the constants $\Lambda_j$ are 
given by \eqref{Lambdaj}).
Then, by multiplying the Wigner equation \eqref{WE} by $1$, $p/m$ and $p^2/2m$, respectively, and integrating both sides with respect to $p$,
we obtain the following system of Euler-like equations:
\begin{equation}
\label{Euler}
\left\{
\begin{aligned}
&\frac{\pt N}{\pt t} + \frac{\pt J}{\pt x}  =  0,
\\[8pt]
&\frac{\pt J}{\pt t}  + \frac{\pt  \cJ_J}{\pt x} + \frac{1}{m}\,V'N= \frac{\hbar\Lambda_1}{m\tau}\,N,
\\[8pt]
&\frac{\pt E}{\pt t}  + \frac{\pt \cJ_E}{\pt x}  + V' J
 = \frac{\hbar\Lambda_1}{m\tau} \,J - \frac{\hbar^2 \Lambda_2}{m \tau}\,N,
\end{aligned}
\right.
\end{equation}
where
$$
\cJ_J(x,t) = \frac{1}{m^2} \int_\mR p^2 w(x,p,t)\,dp = \frac{1}{m} E(x,t)
$$
and
$$
 \cJ_E (x,t) = \frac{1}{2m^2} \int_\mR p^3 w(x,p,t)\,dp
$$
are the currents associated to $J$ and $E$, respectively.
As usual, this system contains the extra unknown $\cJ_E$ (but also $\cJ_J$ would be an unknown in higher spatial dimensions) and needs to be closed by 
making suitable assumptions (see e.g.\ Refs.\ \cite{ChapterKP,Jungel,Romano} and references therein).
\par
The right-hand sides of Eq.\ \eqref{Euler} are due to decoherence collisions. 
We can notice that the terms depending on $\Lambda_1$ are due to the momentum injection from the environment (see the discussion in the first part 
of Sec.\ \ref{S3}), while the term depending on $\Lambda_2$ (which is a positive constant, as it is apparent from \eqref{Lambdaj}) 
represents energy dissipation in the environment.
\section{Large-time asymptotics}
\label{S5}
As $t\to +\infty$, the solution $w(x,p,t)$ to the Wigner equation \eqref{WE} tends to be completely smoothed out to a constant value.
Correspondingly, within the density matrix formalism,  the coherence length associated to $\rho$, i.e.\ the decay of $\rho(X,Y,t)$ along the correlation 
coordinate $X-Y$, tends to vanish.
This unphysical behaviour was already pointed out by Joos and Zeh \cite{JZ}.
\par
Inspired by the approach adopted in Ref.~\cite{JZ}, rather than embarking in a general analysis, we shall discuss the issue 
of large-time asymptotics by performing numerical simulation in a very simple (but physically meaningful) situation, that is the case of a gaussian distribution. 
\par
Let us work within the Wigner-Fokker-Planck approximation \eqref{WFP}, and assume that the potential is harmonic, namely
$$
  V(x) = \frac{\kappa}{2}\,x^2,
$$
with $\kappa \geq 0$.
Recalling Eq.\  \eqref{Thetapprox}, the resulting equation is
\begin{equation}
\label{WFP2}
  \frac{\pt w}{\pt t} + \frac{p}{m} \frac{\pt w}{\pt x} -\kappa \frac{\pt w}{\pt p} =   \frac{\Lambda_0}{\tau} \frac{\pt^2 w}{\pt p^2},
\end{equation}
where for simplicity we have set 
$$
  \Lambda_0 = \hbar^2\Lambda_2.
$$
It is readily seen that \eqref{WFP} admits solutions of the form
\begin{equation}
\label{GWF}
  w(x,p,t) = \e^{-[A(t)\,p^2 + B(t)\,px + C(t)\,x^2 + D(t)]} ,
\end{equation}
where $1/\sqrt{2A(t)}$ is the momentum spread (and, therefore,  $\hbar\sqrt{2A(t)}$ is the coherence-length spread, according to the discussion
closing Section \ref{S2}), $B(t)$ is a covariance parameter, $1/\sqrt{2C(t)}$ is the position spread and $D(t)$ is a normalization parameter.
It is to be noticed that the corresponding density matrix still has a gaussian form, which is exactly the one considered by Joos and Zeh. 
The substitution of \eqref{GWF} into the Wigner-Fokker-Plank equation \eqref{WFP} leads straightforwardly to the following system of ODEs for the
unknown functions  $A(t)$, $B(t)$, $C(t)$ and $D(t)$:
\begin{equation}
\label{EDO1}
\left\{
\begin{aligned}
  &\dot A = -\frac{1}{m}\,B - \frac{4\Lambda_0}{\tau}\,A^2,
\\[6pt]
  &\dot B = -\frac{2}{m}\,C - \frac{4\Lambda_0}{\tau}\,A B + 2\kappa\,A,
\\[6pt]
 &\dot C = -\frac{\Lambda_0}{\tau}\,B^2 + \kappa\,B,
\\[6pt]
&\dot D =  \frac{2\Lambda_0}{\tau}\, A.
\end{aligned}
\right.
\end{equation}
This system for $A$, $B$ and $C$ (which is decoupled from the equation for $D$) possesses the unique, asymptotically stable, equilibrium point 
$(A,B,C) = (0,0,0)$. 
This means that, as expected, the Wigner function is completely smoothed out towards a constant value (which is of course $0$).
Correspondingly, the coherence length goes to zero. 
The model is therefore not satisfactory for large times, since, as remarked by Joos and Zeh \cite{JZ}, the coherence must be maintained at least at the 
length-scale  of the thermal De Broglie wavelength 
\begin{equation}
\label{TDB}
  \lambda_\mathrm{th} = \frac{\hbar}{\sqrt{2mk_BT}},
\end{equation}
where $T$ is the temperature of the environment particle bath, and $k_B$ is the Boltzmann constant.
\par
By looking at the equilibrium conditions for system \eqref{EDO1} we can guess that the addition to the first equation 
of a linear term in $A$, with positive coefficient, is able to shift the equilibrium from $A = 0$ to a positive value. 
How such a term could arise from Eq.\ \eqref{WFP2}?
If we want to preserve the gaussian form \eqref{GWF} of the solution, we see that there are not many more possibilities than
adding a derivative of $w$ with respect to $p$ and multiply it by $p$.
We realised that this is provided by a ``quantum friction'' term proposed by Caldeira and Legget \cite{CL}.
In fact, for high temperatures and in the density matrix formalism, this term appears at the right-hand side of the von Neumann equation \eqref{VNEdeco}
as 
$$
 i\hbar\frac{\eta}{2} (X-Y)\left(\frac{\pt \rho}{\pt X} - \frac{\pt \rho}{\pt Y}\right) ,
$$
where $\eta \geq 0$ is a ``friction'' coefficient \cite{CL,DH}.
Translating this term into the Wigner formalism, and adding it to the Wigner equation \eqref{WFP2},  we obtain 
\begin{equation}
\label{WFP3}
  \frac{\pt w}{\pt t} + \frac{p}{m} \frac{\pt w}{\pt x} -\kappa \frac{\pt w}{\pt p} =   \frac{\Lambda_0}{\tau} \frac{\pt^2 w}{\pt p^2}
   + \eta \, \frac{\pt }{\pt p}(p w) ,
\end{equation}
which has exactly the needed form.
Substituting \eqref{GWF} in \eqref{WFP3} yields the new system of ODEs
\begin{equation}
\label{EDO2}
\left\{
\begin{aligned}
  &\dot A = -\frac{1}{m}\,B - \frac{4\Lambda_0}{\tau}\,A^2 + 2\eta\,A
\\[6pt]
  &\dot B = -\frac{2}{m}\,C - \frac{4\Lambda_0}{\tau}\,A B + 2\kappa\,A + \eta\,B,
\\[6pt]
 &\dot C = -\frac{\Lambda_0}{\tau}\,B^2 + \kappa\,B,
\\[6pt]
&\dot D =  \frac{2\Lambda_0}{\tau}\, A - \eta,
\end{aligned}
\right.
\end{equation}
possessing the asymptotically stable equilibrium point 
\begin{equation}
\label{equil}
  (A_0,B_0,C_0) = \left(\frac{\tau\eta}{2\Lambda_0}, 0,  \frac{m\tau\kappa\eta}{2\Lambda_0}\right).
\end{equation}
Note that the asymptotic coherence length is
$$
   \hbar\sqrt{2A_0} = \hbar\,\sqrt{\frac{\tau\eta}{\Lambda_0}} = \frac{\hbar}{\sqrt{mk_BT}},
$$
where the last equality holds if one takes the relation 
$$
  \Lambda_0 = \tau m\eta k_BT,
$$ 
as it is done, e.g., in Ref.\ \cite{DH}.
Hence, we obtain that the asymptotic coherence length is
of the order of the thermal De Broglie wavelength \eqref{TDB}, exactly as physically expected.
We also note that the simultaneous presence of the friction and of the harmonic potential stabilises the position spread towards  the 
asymptotic value
$$
  \frac{1}{\sqrt{2C_0}} = \sqrt{\frac{\Lambda_0}{2\tau\kappa \eta}} = \sqrt{\frac{k_BT}{\kappa}}
$$
(where the last equality holds if one takes $\Lambda_0$ as above).
\par
\begin{figure}[htb]
\begin{center}
\includegraphics[scale=0.45,clip= true]{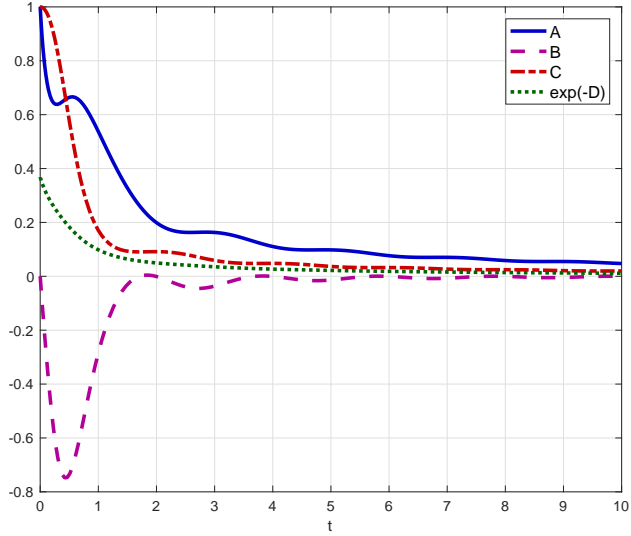} 
%\begin{spacing}{1.0}
\caption{(Colors online.) Evolution of the parameters $A$ (continuous blue line), $B$ (purple dashed line) and $C$ (red dot-dashed line)
of the Wigner function \eqref{GWF} in absence of friction ($\eta = 0$).
The overall normalisation coefficient $\exp(-D)$ (dotted green line) is also shown.
We assume to work in arbitrary units in which $m = 0.4$, $\tau = 1$, $\Lambda_0 = 1$ and $\kappa = 1$.
\label{figura1}}
%\end{spacing}
\end{center}
\end{figure}
\begin{figure}[htb]
\begin{center}
\includegraphics[scale=0.45,clip= true]{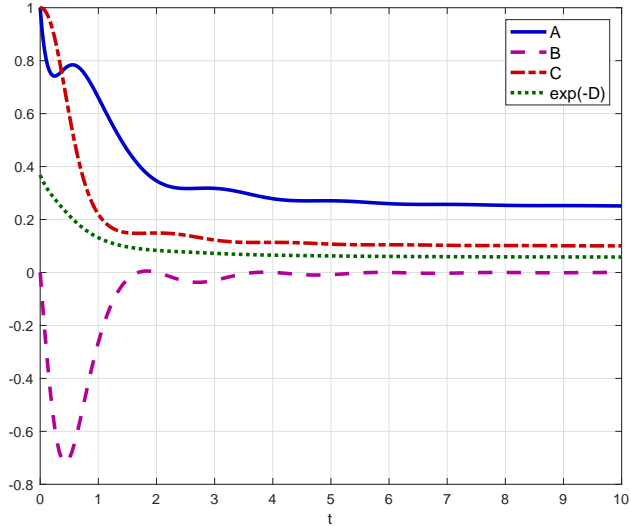} 
%\begin{spacing}{1.0}
\caption{
The same as in Figure \ref{figura1} but with the addition of the friction. 
The values of the parameters are $m = 0.4$, $\tau = 1$, $\Lambda_0 = 1$, $\kappa = 1$ and $\eta = 0.5$.
\label{figura2}}
%\end{spacing}
\end{center}
\end{figure}
\begin{figure}[htb]
\begin{center}
\includegraphics[scale=0.45,clip= true]{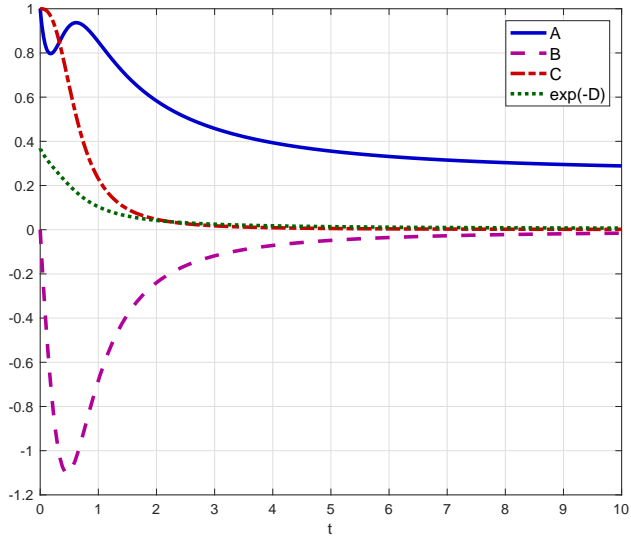} 
%\begin{spacing}{1.0}
\caption{The same as in Figure \ref{figura2} but with the harmonic potential removed. 
The values of the parameters are $m = 0.4$, $\tau = 1$, $\Lambda_0 = 1$, $\kappa = 0$ and $\eta = 0.5$.
\label{figura3}}
%\end{spacing}
\end{center}
\end{figure}

In Figures \ref{figura1}--\ref{figura3} we show some solutions to system \eqref{EDO2}.
In Fig.\ \ref{figura1} we set $\eta = 0$ and we can see that in the absence of friction both $A(t)$ and $C(t)$ approach zero as $t\to\infty$.
This means that the Wigner function becomes infinitely spread out in both momentum and position, and tends to zero everywhere ($D(t) \to +\infty$).
When a friction is added (Fig.\ \ref{figura2}), both the momentum and the position spread  stabilise to their asymptotic values \eqref{equil}.
In this case, the Wigner function does not vanish, since $D(t)$ tends to an asymptotic positive value.
When the harmonic trap is switched off by putting $\kappa = 0$, we can see that friction is able to stabilise the momentum spread but 
not the position spread (Fig.\ \ref{figura3}). Consequently, the Wigner function becomes completely spread out in the $x$ direction and, 
as in the case of Fig.\ \ref{figura1}, tends to vanish ($D(t) \to +\infty$).
\par
In the figures we use arbitrary units, where, in particular, $\tau = 1$.
The actual decoherence time depends of course on the considered system (namely, the size and mass of the particle, the scattering properties and the 
temperature of the environment, and so on). We refer the reader to the accurate discussion contained in Ref.~\cite{JZ}.
\par
The dissipative character of the new term is clearly seen by computing its moments:
$$
%\begin{aligned}
  \int_\mR \frac{\pt}{\pt p}\,(p w)\,dp = 0,
\qquad
  \frac{1}{m} \int_\mR p \frac{\pt}{\pt p}\,(p w)\,dp = -J,
\qquad
 \frac{1}{2m}\int_\mR p^2\frac{\pt}{\pt p}\,(p w)\,dp = -2E.
%\end{aligned}
$$
These bring dissipative contributions to the Euler system, which takes the new form 
\begin{equation}
\label{Euler2}
\left\{
\begin{aligned}
&\frac{\pt N}{\pt t} + \frac{\pt J}{\pt x}  =  0,
\\[8pt]
&\frac{\pt J}{\pt t}  + \frac{\pt \cJ_J}{\pt x}  + \frac{1}{m}\,V'N= \frac{\hbar\Lambda_1}{m\tau}\,N - \eta J,
\\[8pt]
&\frac{\pt E}{\pt t}  + \frac{\pt \cJ_E}{\pt x}  + V' J
 = \frac{\hbar\Lambda_1}{m\tau} \,J - \frac{\hbar^2 \Lambda_2}{m \tau}\,N - 2\eta E.
\end{aligned}
\right.
\end{equation}
\section{Conclusions}
\label{S6}
In this paper we have seen how Wigner equation can be endowed with terms describing a dynamical decoherence mechanism.
This is not a novelty, of course, but, as far as we know, this is the first time that the decoherence term has a fairly general form, coming from 
basic quantum mechanics.
In particular, we started from the single-collision decoherence model derived in Ref.\ \cite{AHN}, which describes the decoherence of a  ``heavy'' particle  
as a consequence of the collision with a much lighter one.
By assuming the heavy particle to undergo multiple random collisions in an environment of light particles, we (formally) derived the Wigner equation 
\eqref{WE}. 
The latter admits two contributions from the collisions with the environment: a Hamiltonian part, represented by the function $\Gamma$,
and a true decoherence part, represented by the function $\Lambda$, which is nothing but the inverse Fourier transform of the convolution kernel 
$\gamma$ appearing in the right-hand side of Eq.\ \eqref{WE}.
This picture allows for the interesting interpretation that decoherence smooths out the oscillations of the Wigner function, due to quantum interference,
so that the Wigner function tends to a classical distribution in phase-space.
 
Then, we have seen that when $\Lambda$ assumes particular forms, our model reduces to existing decoherence models.
In particular, the largely-used Wigner-Fokker-Planck equation \eqref{WFP} corresponds to the quadratic approximation of $\Lambda$. 
Moreover, when $1-\Lambda$ is assumed to be a decaying exponential function, our model shows analogies with the Jacoboni-Bordone model \cite{JB}, 
in which the exponential decay of the coherence length is embedded {\em ab initio} in the definition of the Wigner function.
Our analysis, however, allows us to deduce some general features of decoherence (or, at least, of this kind of decoherence), as for example its 
effects on the dynamics of the macroscopic quantities $N$, $J$ and $E$, i.e.\ the number, current and energy spatial densities 
(see Eq.\ \eqref{Euler}).

A big issue, already addressed in the classical paper by Joos and Zeh \cite{JZ} is the long time behaviour of decoherence.
In Section \ref{S5} we have considered the special case of a gaussian Wigner function, for which the Wigner equation \eqref{WE}, in the particular
form \eqref{WFP2}, comes down to an equivalent system if ODEs.
In this way we realized that the addition of a Caldeira-Legget quantum friction term \cite{CL} produces a physically meaningful
behaviour in the long run, since the momentum spread of the particle is stabilised to an asymptotic value. 
Equivalently, the coherence length, reaches a corresponding asymptotic value.

The addition of the quantum friction fixes the issue of the long-time behaviour, yet it is not completely satisfactory.
In fact, as remarked by Arnold et al.\ \cite{ALMS,Arnold}, the friction + diffusion term 
(i.e.\ the right-hand side of Eq.\ \eqref{WFP3}) is not quantum mechanically ``correct'' (unless $\eta = 0$), since it does not satisfy  the
Lindblad condition, assuring the complete positivity of the evolution \cite{Da}.
We believe, however, that our analysis indicates the right direction to search for a model that is compatible with the fundamental 
laws of quantum mechanics and keeps its validity for asymptotically long times.

\section*{Acknowledgments}
L.\ B.\  acknowledges support from the Italian-French project PICS (Projet International de Coop\'eration Scientifique) 
``MANUS - Modelling and Numerics for Spintronics and Graphene'' (Ref. PICS07373). 
\par
E.\ G.\ acknowledges support from INdAM-GNFM (Italian National Group for Mathematical Physics).
%(Junior project (2017) {\em Hydrodynamic models for quantum and spinorial dynamics}.
\par
Part of this work was accomplished during E.\ G.\  staying at Paul Sabatier University in Toulouse with a grant MAE-2018 from the French
{\it Minist\`ere de l'Europe et des Affaires \'etrang\`eres}.
\par
The authors wish to express their gratitude to Prof.\ Claudia Negulescu for her valuable support and for many stimulating discussions.
%
%

% for those who wish not to use a .bib file:

\setlength{\baselineskip}{12pt}

\end{document}